\documentclass[aps,prl,twocolumn,showpacs,superscriptaddress,letterpaper]{revtex4}
\usepackage{amsmath,amssymb,graphicx,epsfig,color,bbm}
\usepackage[pass]{geometry}

\def\nn{\nonumber}
\def\llangle{\langle\!\langle}
\def\rrangle{\rangle\!\rangle}

\begin{document}

\title{Limit cycle phase in driven-dissipative spin systems}

\author{Ching-Kit Chan}
\affiliation{ITAMP, Harvard-Smithsonian Center for Astrophysics, Cambridge, Massachusetts 02138, USA}
\affiliation{Department of Physics, Harvard University, Cambridge, Massachusetts 02138, USA}
\author{Tony E. Lee}
\affiliation{ITAMP, Harvard-Smithsonian Center for Astrophysics, Cambridge, Massachusetts 02138, USA}
\affiliation{Department of Physics, Harvard University, Cambridge, Massachusetts 02138, USA}
\author{Sarang Gopalakrishnan}
\affiliation{Department of Physics, Harvard University, Cambridge, Massachusetts 02138, USA}

\date{\today}

\begin{abstract}
We explore the phase diagram of interacting spin-$1/2$ systems in the presence of anisotropic interactions, spontaneous decay and driving. We find a rich phase diagram featuring a limit cycle phase in which the magnetization oscillates in time. We analyze the spatio-temporal fluctuations of this limit cycle phase at the Gaussian level, and show that spatial fluctuations lead to quasi-long-range limit cycle ordering for dimension $d = 2$. This result can be interpreted in terms of a spatio-temporal Goldstone mode corresponding to phase fluctuations of the limit cycle. We also demonstrate that the limit-cycle phase exhibits an asymmetric power spectrum measurable in fluorescence experiments.
\end{abstract}


\maketitle


A quantum system that is coherently driven and connected to a heat bath eventually reaches a steady state; when the system is macroscopic, this steady state can be ordered in the sense of spontaneously breaking a symmetry \cite{altland10,sachdev11}. The patterns of steady-state ordering are, in general, qualitatively different from those of the equilibrium phase diagrams: for instance, the interplay between coherent and dissipative dynamics can stabilize staggered phases that are absent in equilibrium \cite{lee13}, and by engineering the dissipative terms one can optically pump a many-body system into a pure state \cite{diehl08,verstraete09}. The forms of steady-state order \cite{grinstein90,mitra06,vogl12,lee12,leboite13,sieberer14,zou14,honing14,marcuzzi14,weimer14} that have been investigated to date mostly fit into equilibrium \emph{paradigms}, such as the Landau paradigm of spontaneous symmetry breaking or the paradigm of topological order. However, the departure from equilibrium allows one to realize novel types of order---specifically, limit cycles (LC) \cite{strogatz94,rodrigues07,lee11,qian12,jin14} which break time-translation invariance---that have no obvious equilibrium counterpart. Although previous works have predicted a LC phase within mean-field theory \cite{lee11,qian12,jin14}, it is an open question whether such a phase really exists or is an artifact of mean-field theory.

In this work, we show that the LC phase exists with long-range order in three and higher dimensions and quasi-long-range order in two dimensions. We study a paradigm model of interacting spins---the anisotropic spin-1/2 Heisenberg model in a transverse field---and find a regime where it exhibits a limit-cycle phase. We discuss the origin of the LC and study the effects of quantum fluctuations to go beyond previous mean-field works. We show by explicit, microscopic calculation that the spontaneous breaking of the continuous time-translation symmetry is reflected in the presence of a gapless ``Goldstone'' mode, and that this gapless mode prevents global time-translation-symmetry-breaking in one or two dimensions. We then discuss this LC at a more phenomenological level, noting its unusual implications, e.g., that the temporal ordering of the LC phase gives rise to an \textit{asymmetric} dynamical power spectrum of emitted photons. These predictions are straightforward to test in experiments with trapped ions \cite{molmer99,islam11,britton12} or Rydberg atoms \cite{bouchoule02,lee13,glaetzle14}.

We want to contrast our paper with recent works on time crystals which also spontaneously break time-translational symmetry \cite{wilczek12,shapere12,li12,wilczek13,bruno13,volovik13}. A LC is the nonequilibrium steady state of a master equation, whereas a time crystal is the ground state of a Hamiltonian; thus, LCs are unaffected by the no-go theorems concerning equilibrium time crystals \cite{bruno13}.

\begin{figure}[b]
\begin{center}
\includegraphics[angle=0, width=1\columnwidth]{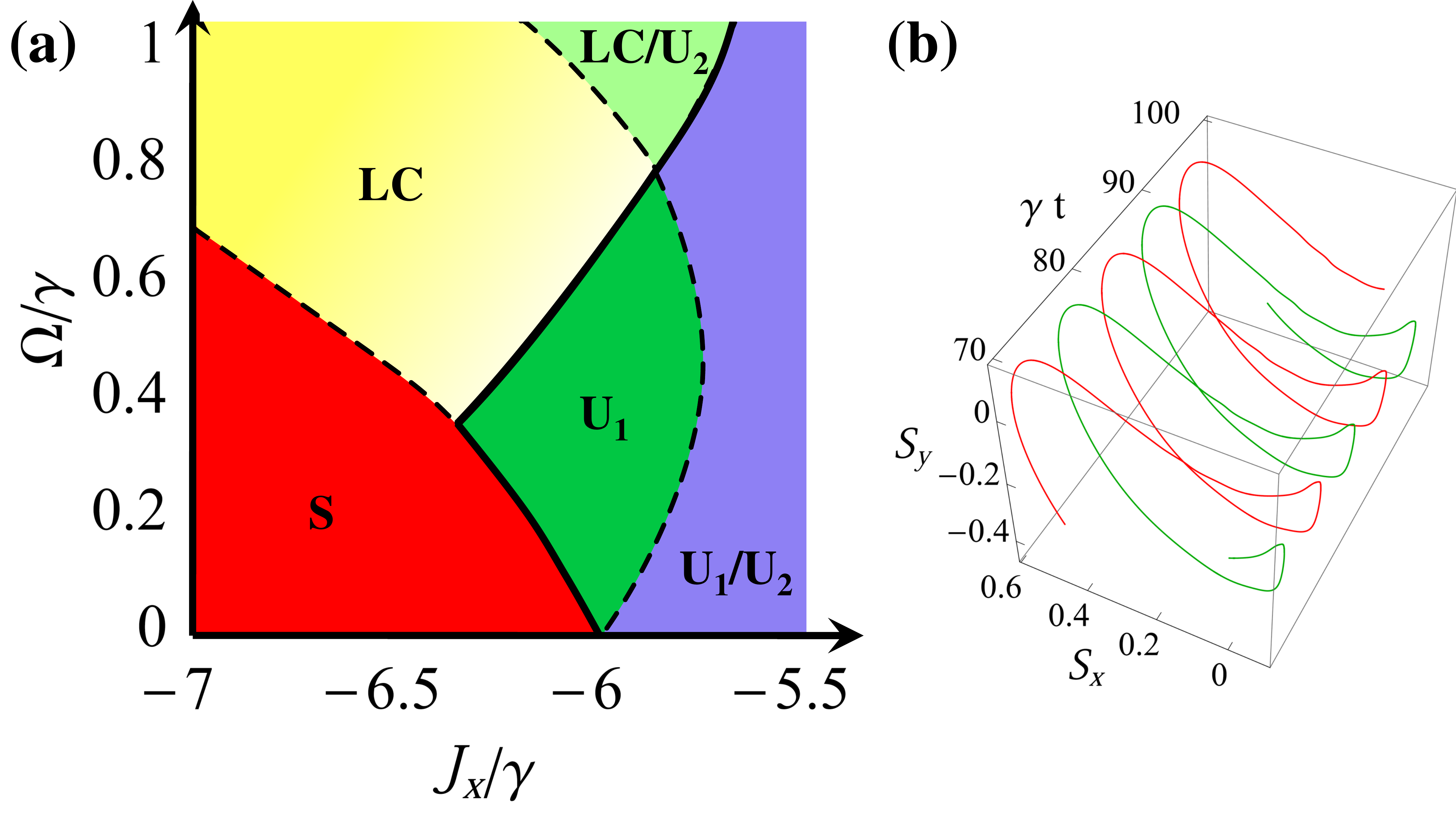}
\caption{(Color online) (a) Mean-field phase diagram of the driven-dissipative spin system for $\left(J_y,J_z\right)/\gamma=\left(6,2\right)$, highlighting the limit cycle and its vicinity. The nonequilibrium spin system has different phases: uniform ($\text{U}_1$), bistable ($\text{U}_1/\text{U}_2$), staggered ($\text{S}$), and LC phases. LC/$\text{U}_2$ denotes phase coexistence. In the LC phase, the spin system undergoes self-sustaining collective oscillations in time. The solid and dashed lines represent continuous and discontinuous phase transitions, respectively. (b) An example trajectory of the time-dependent LC at $(\Omega,J_x,J_y,J_z)/\gamma=(1,-7,6,2)$ that breaks both the time-translational and sublattice symmetry. }
\label{fig_phase_diagram}
\end{center}
\end{figure}

\textit{Model and phase diagram}.---Consider a $d$-dimensional square lattice of spins with nearest-neighbor exchange interaction. Each spin is resonantly driven by a laser and experiences spontaneous decay that incoherently flips the spin from up to down. The resultant many-body dynamics is governed by a master equation for the system's density matrix $\rho$:
\begin{eqnarray}
\label{eq_ME}
\dot \rho &=& -i \left[ H_J+H_\Omega, \rho \right] - \gamma \sum_i \left[ \frac{1}{2} \left\{\sigma_i^+ \sigma_i^-,\rho \right\}-\sigma_i^- \rho \sigma_i^+ \right], \nn \\
H_J &=& \frac{1}{2d}\sum_{\langle i,j\rangle,\alpha} J_\alpha \sigma_i^\alpha \sigma_j^\alpha, \ \ \ \ \
H_\Omega =\frac{\Omega}{2} \sum_i \sigma_i^x,
\end{eqnarray}
where $\sigma_i^\alpha$ is the $\alpha$ component Pauli matrix for spin $i$. $H_J$ and $H_\Omega$ describe the Heisenberg interaction and coherent driving, respectively. $\gamma$ is the spontaneous decay rate, experimentally introduced via optical pumping. The exchange interaction can be realized in trapped ions via motional sidebands \cite{molmer99,islam11,britton12}, and Rydberg systems based on optical adiabatic elimination \cite{bouchoule02,lee13,glaetzle14}. $J$ and $\gamma$ can be engineered to be on the order of $1~\text{kHz}$ with trapped ions or $100~\text{kHz}$ with Rydberg atoms. In the following, we focus on the parameter regime $J \sim \Omega \sim \gamma$.

The resulting dynamics can be understood in terms of the interplay between spontaneous decay, drive, and spin interaction. When there is no interaction, a spin is driven around the $x$ axis and equilibrates in the lower $yz$ plane (so that $\langle\sigma^x\rangle=0, \langle\sigma^y\rangle \neq 0, \langle\sigma^z\rangle < 0$). In the presence of interaction, each spin precesses about an effective field ($\sim \sum_\alpha J_\alpha \langle{\sigma^\alpha }\rangle  \hat\alpha$) due to neighboring spins. This effective field is established self-consistently, leading to the possibility of ordered states. When the coupling is anisotropic ($J_x \sim -J_y$), the effective field becomes almost perpendicular to the spin direction, and the precession effect is stronger. Therefore, we expect a richer phase diagram in the anisotropic coupling regime. Note that this system does not possess any $Z_2$ symmetry (like $\sigma_i^x,\sigma_i^y \rightarrow -\sigma_i^x,-\sigma_i^y$) due to the presence of the drive. The only symmetries of the master equation are a continuous symmetry under time translation and a discrete symmetry under spatial (lattice) translation. We restrict our analysis to bipartite lattices; thus, the mean field on sublattice A is due to the magnetization on sublattice B and vice versa.

\begin{figure}[t]
\begin{center}
\includegraphics[angle=0, width=0.48\textwidth]{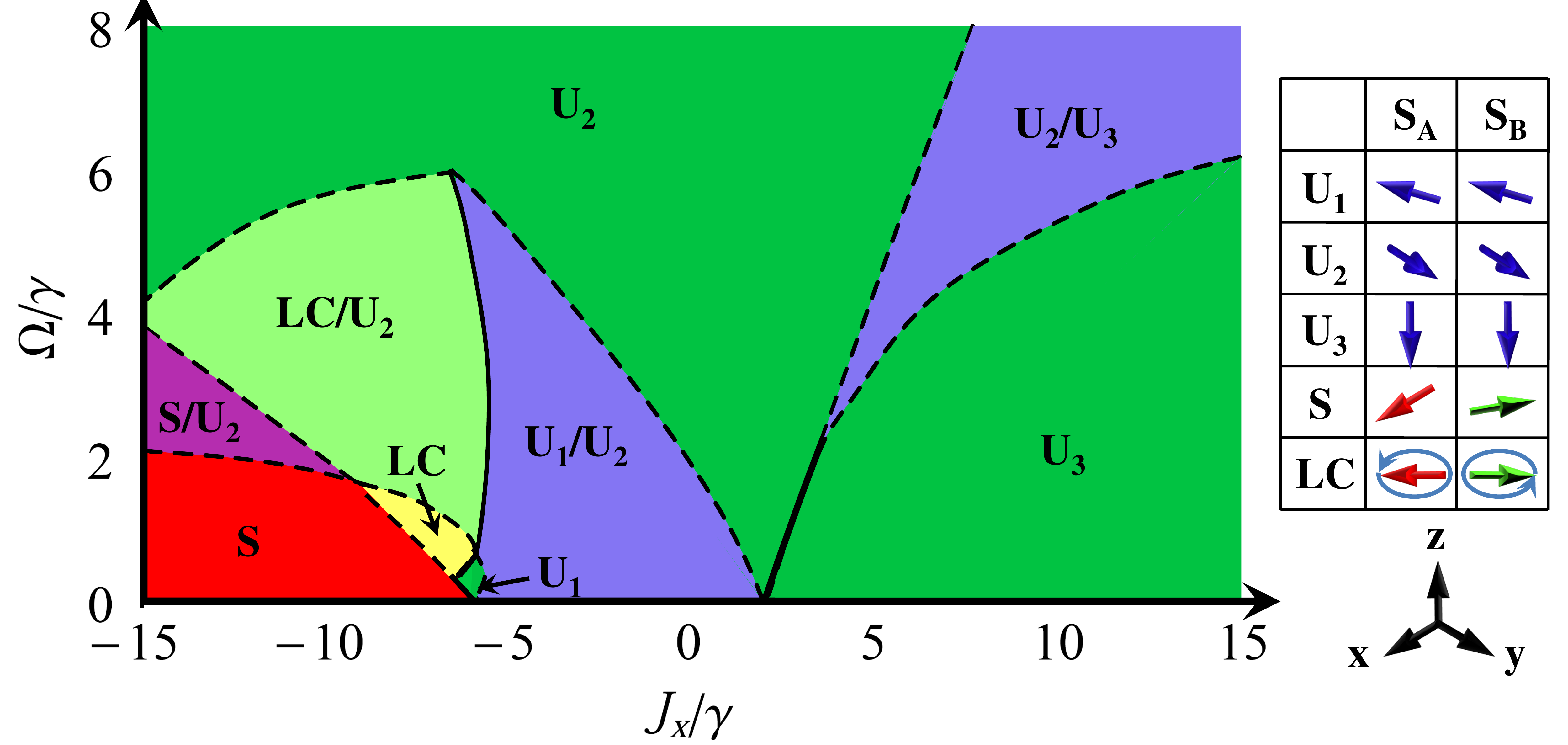}
\caption{Zoomed-out phase diagram showing different phases for $\left(J_y,J_z\right)/\gamma=\left(6,2\right)$.}
\label{fig_phase_diagram_zoomout}
\end{center}
\end{figure}

\begin{figure}[t]
\begin{center}
\includegraphics[angle=0, width=1\columnwidth]{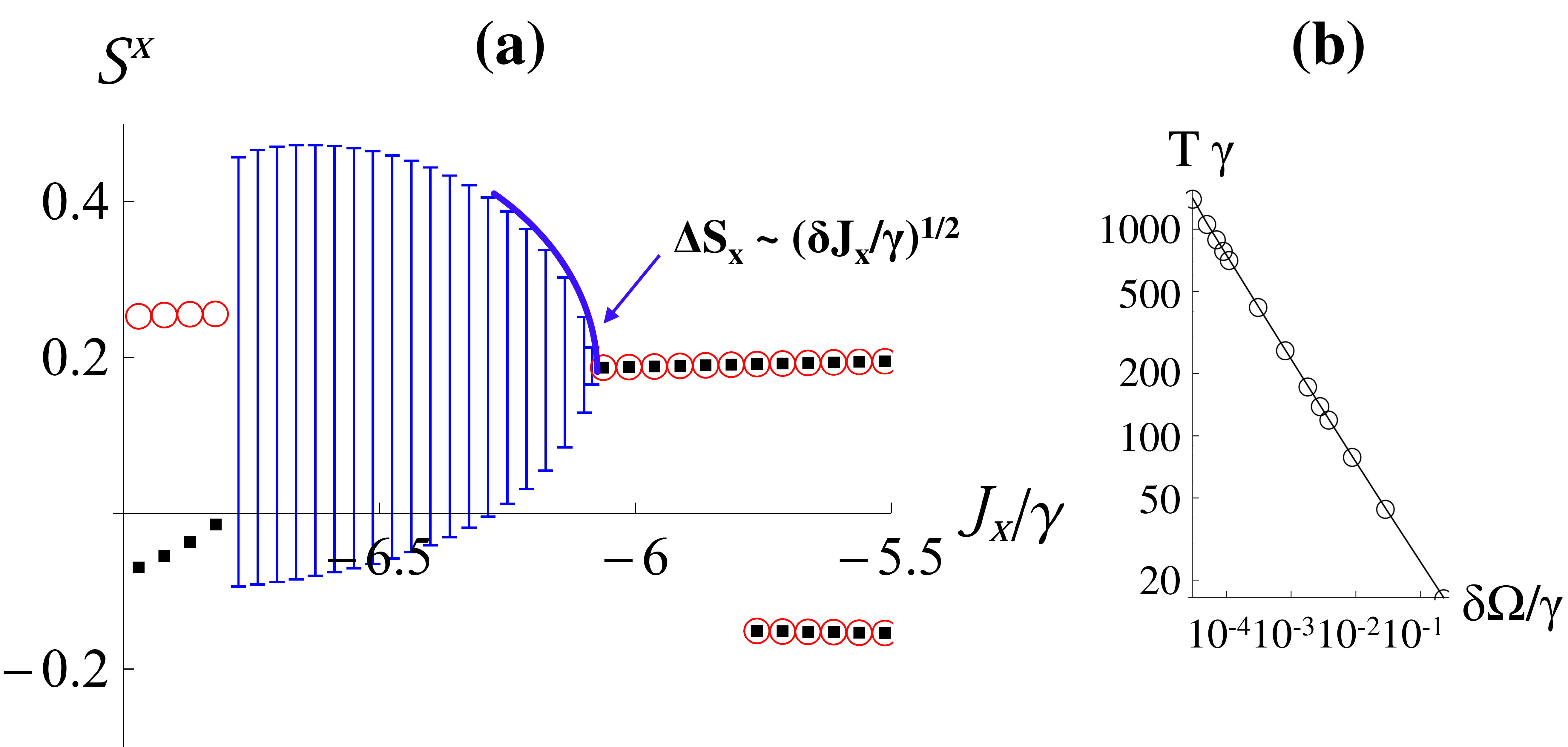}
\caption{(Color online) (a) Phase transitions of the LC phase as a function $J_x$ at $(\Omega,J_y,J_z)/\gamma=(0.6,6,2)$. The sublattice spin order parameter $S^x$ develops bistability, Hopf bifurcation and sublattice order. The LC phase has a continuous and discontinuous phase transition to the $\text{U}_1$ and S phase, respectively. Vertical bars represent the amplitudes of the sublattice spin oscillations. Circles and solid squares denote the two sublattice stationary states. The LC amplitude shows a scaling of $\left( \delta J_x/\gamma\right)^{1/2}$ near the continuous LC-to-$\text{U}_1$ transition. (b) LC period $T$ as we vary $\Omega$ for fixed $(J_x,J_y,J_z)/\gamma=(-6.37,6,2)$. The divergence of period scales as $T \sim \left( \delta \Omega/\gamma\right)^{-1/2}$ as we approach the tricritical point at $(J_{x,c},\Omega_c)/\gamma \approx (-6.37,0.37)$.}
\label{fig_steady_state}
\end{center}
\end{figure}

A typical scan of the steady-state phase diagram is shown in Figs.~\ref{fig_phase_diagram}(a) and~\ref{fig_phase_diagram_zoomout}. The calculation is based on sublattice mean-field theory and linear stability analysis to obtain the stable fixed points of Eq.~(\ref{eq_ME}). There are (i) three qualitatively different uniform steady states ($\text{U}_{1,2,3}$) that do not break any symmetry and predominate at large drive and/or isotropic couplings; (ii) a staggered phase ($\text{S}$), which is time-independent but breaks the sublattice symmetry, and dominates for strongly anisotropic coupling; and (iii) a LC phase, which breaks both sublattice and time-translation symmetries.

The uniform phases are related to the paramagnetic ($\text{U}_3$) and the two ferromagnetic ($\text{U}_{1,2}$) steady states of the undriven system; in the presence of the drive, $Z_2$ symmetry is explicitly broken so these phases are no longer distinct in terms of symmetry. The staggered phase is a descendant of the antiferromagnet; since it breaks lattice translation symmetry, it is quite different from the uniform phases; thus the uniform-staggered transition is second-order. Finally, the LC phase, which has no analog in the undriven system, emerges near the uniform-staggered phase boundary in the presence of a drive. It is intuitive that the LC arises in this regime, since even in the undriven system the relative orientation between the two sublattices is ``soft'' (i.e., highly susceptible) near the transition between the ferromagnet ($\text{U}_1$) and the antiferromagnet ($\text{S}$).

\emph{Limit cycle: mean-field analysis}.---In the LC phase, the magnetizations on the two sublattices oscillate with a relative phase of $\pi$ [Fig.~\ref{fig_phase_diagram}(b)]. Thus, the LC breaks both spatial and time translation symmetry.

Figure~\ref{fig_steady_state} (a) illustrates the properties of the LC phase transitions. The plot shows the mean-field amplitude of $S^x =\langle \sigma^x \rangle$ as we vary $J_x$ with fixed $\Omega$, $J_y$ and $J_z$. As $J_x$ decreases, the spin system goes from a bistable phase ($\text{U}_1/\text{U}_2$) to a uniform ($\text{U}_1$) phase, and then exhibits a supercritical Hopf bifurcation \cite{strogatz94,cross93} to the LC phase at the critical point of $J_{x,c}\approx -6.08\gamma$. A further decrease of $J_x$ to $\approx -6.78\gamma$ renders a first-order transition from LC phase to the staggered phase. Note that in this LC phase, there are no stable fixed points. The amplitude of the LC scales as $\left( \delta J_x/\gamma\right)^{1/2}$ near the second-order phase transition from LC to $\text{U}_1$, with $\delta J_x = J_{x,c}-J_x$ \cite{nayfeh11}.

The intersection of continuous LC-$\text{U}_1$ and discontinuous LC-S phase boundaries defines a LC tricritical point [Fig.~\ref{fig_phase_diagram}(a)]. Interestingly, the LC period diverges near this tricritical point [Fig.~\ref{fig_steady_state}(b)]. The reason stems from the imaginary part of stability eigenvalues that determine the frequency of LC \cite{nayfeh11}. Along the LC-$\text{U}_1$ boundary, the two stability eigenvalues are purely imaginary complex conjugates, corresponding to the Hopf bifurcation, while on the S-$\text{U}_1$ boundary, they both vanish due to the sublattice symmetry. Therefore, the LC near their intersection has a divergent period \cite{hu13}.

\emph{Gaussian fluctuations}.---We now go beyond mean-field theory and explore Gaussian fluctuations about the mean-field states. Because the mean-field LC spontaneously breaks time-translation invariance, fluctuations about it obey an inhomogeneous Floquet equation \cite{yakubovich75,slane11}. The correlation functions can be expressed as (see Supplemental Material \cite{supple}):
\begin{eqnarray}
\label{eq_corr_equation}
\boldsymbol{\dot C}(k,t) = \boldsymbol A(k,t) \boldsymbol  C(k,t) +\boldsymbol C(k,t) \boldsymbol  A^T (k,t) + \boldsymbol  D(k,t),
\end{eqnarray}
where $\boldsymbol C(k,t)$ is the Fourier transform of the spatial correlation matrix $C^{\alpha,\beta}_{L,L'}(r,t)= \langle \sigma^\alpha_{L} (0,t) \sigma^\beta_{L'}(r,t)\rangle -\langle \sigma^\alpha_{L} (0,t) \rangle \langle \sigma^\beta_{L'}(r,t)\rangle $ with $L$ and $L'$ denoting sublattices. $\boldsymbol A$ and $\boldsymbol D$ are the drift and the diffusion matrices, respectively \cite{carmichael99}, and are both periodic in time due to dependence on the mean-field variable $\vec S_L(t)$. So, $\boldsymbol C$ is also time-dependent. In contrast, for a stationary phase, the steady state solution for $\boldsymbol C$ does not have any time dependence.

\begin{figure}[t]
\begin{center}
\includegraphics[angle=0, width=1\columnwidth]{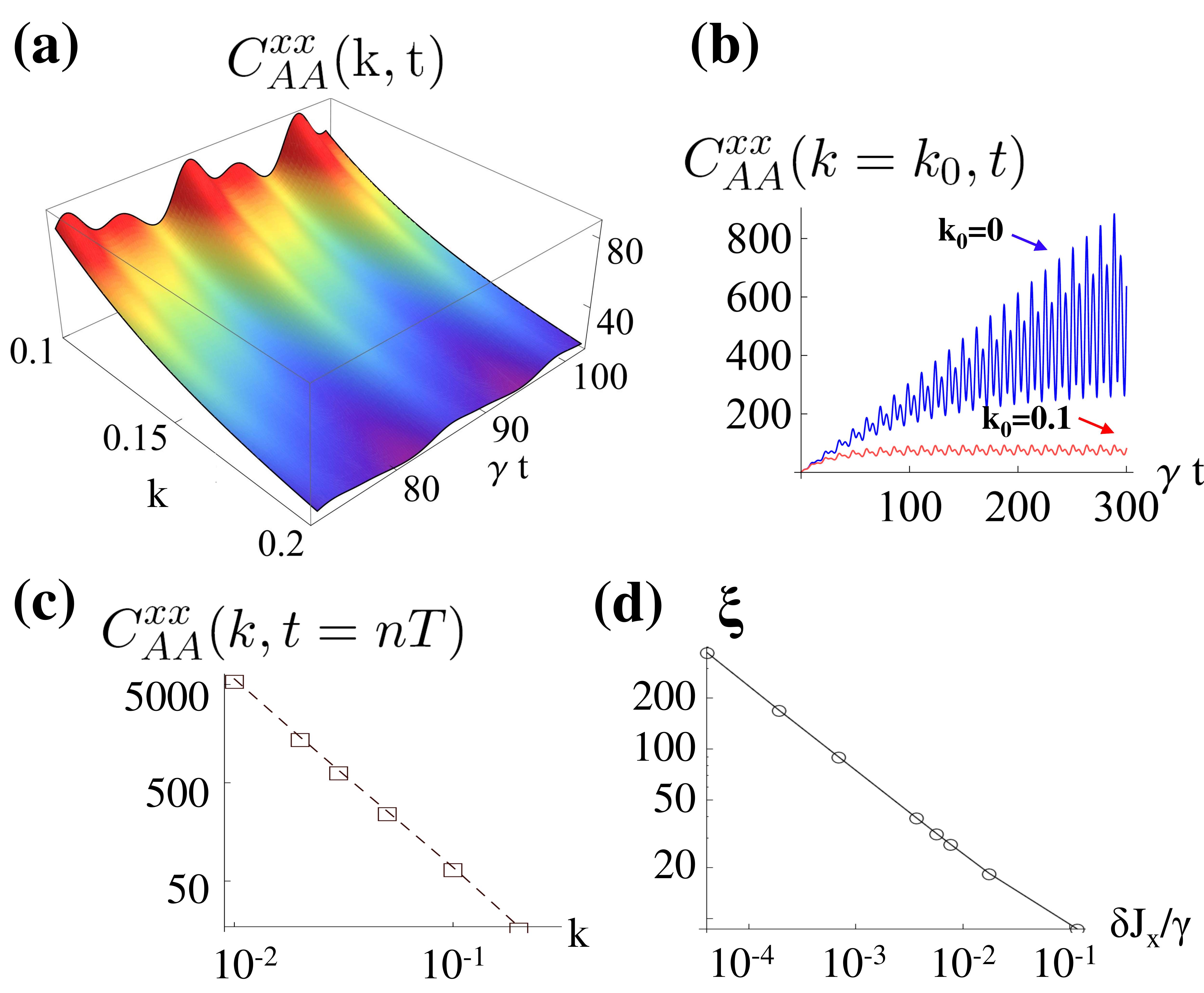}
\caption{(Color online) (a-c) Dynamical spin correlations when $(\Omega,J_x,J_y,J_z)/\gamma=(0.6,-6.09,6,2)$. (a) A typical plot of the sublattice correlation function as a function of momentum and time in the long-wavelength limit. (b) When $k=0$, the correlation diverges linearly with time, while it saturates to finite values for $k \neq 0$. The oscillations have the same period as the LC. (c) The stroboscopic correlation in the long time limit shows a $1/k^2$ dependence, reflecting the gapless excitation of the broken time-translational symmetry of the LC phase. (d) Correlation length $\xi$ scales as $\xi \sim \left( \delta J_x/\gamma\right)^{-1/2}$ near the continuous LC-$\text{U}_1$ phase transition.}
\label{fig_fluctuation}
\end{center}
\end{figure}

In the case of a LC, the correlation function has interesting dynamical behavior due to the time periodicity of $\boldsymbol A$ and $\boldsymbol D$. Figure~\ref{fig_fluctuation}(a) presents a typical plot of the momentum and time dependence of the solution of Eq.~(\ref{eq_corr_equation}) in the LC phase. Figure~\ref{fig_fluctuation}(b) shows that the oscillating correlation function reaches a steady value for finite $k$, but diverges linearly in time at $k=0$. In the long time limit, the stroboscopic correlation goes as $1/k^2$ for small momentum [Fig.~\ref{fig_fluctuation}(c)]. This infrared divergence precludes LC ordering for dimensions $d\leq 2$, and---as we discuss below---can be traced to the Goldstone mode \cite{altland10} generated by spontaneous time-translational symmetry breaking in the LC phase. In $2 < d < 4$, a similar analysis can be used to compute correlation lengths [Fig.~\ref{fig_fluctuation}(d)], and to estimate the size of the fluctuation-dominated region around the onset of a LC via a Ginzburg criterion (see Supplemental Material \cite{supple}). In three dimensions, the critical region has a parameter-space width $\delta J_x \ll 10^{-1} \gamma$, which is of the order of $10~\text{kHz}$ in atomic experiments.

\emph{Phenomenological description}.---The analysis above has the advantage of being fully microscopic, thus enabling comparisons with experiment; however, it does not give a transparent picture of the origin of the Goldstone mode. In the spirit of Landau theory, we now consider a general phenomenological description of a LC, known as its normal form \cite{nayfeh11}. The idea is to expand the order parameter around the time-independent component $\vec S_c$ near the onset of a Hopf bifurcation: $\vec S(t) - \vec S_c = A(t) \vec p + \text{c.c.} $ with the vector $\vec p$ lying on the plane of oscillation. Following standard procedures \cite{nayfeh11}, the mean-field equation can be rewritten near the LC-$\text{U}_1$ boundary as:
\begin{eqnarray}
\label{eq_normal_form}
\frac{d A(\mathbf{x},t)}{dt} & = & \left(\lambda + \frac{2\pi i}{T} + \mathcal{S} \nabla^2 \right) A(\mathbf{x},t) \nonumber \\ && \quad +  \mu \left|A(\mathbf{x},t)\right|^2 A(\mathbf{x}, t) + \zeta(t),
\end{eqnarray}
where $A$, $T$ are the amplitude and period of the LC, respectively. We have augmented the normal form with a stiffness $\mathcal{S}$ against spatial fluctuations, as well as a Langevin noise term \cite{note1}. The stability of the LC requires the coefficients $\lambda \in \Re $ (being linear in $\delta J_\alpha$) and $\mu$ to satisfy $\lambda~ \text{Re }\mu <0$. Importantly, Eq.~(\ref{eq_normal_form}) reveals the continuous $U(1)$ symmetry ($A \rightarrow A e^{i\psi}$) of the LC.

The normal form Eq.~\eqref{eq_normal_form} can be used to derive the lower critical dimension by the following procedure (see Supplemental Material \cite{supple}): working in the rotating frame ($A \rightarrow A e^{2\pi i t/T}$), expanding $A$ in terms of phase fluctuations to obtain a phase-only action, and using this action to compute the correlation function of the phase of the LC. In momentum space, this correlator is $ \langle \theta_\mathbf{k} \theta_{\mathbf{-k}} \rangle \sim 1 / k^2$, and thus its inverse Fourier transform in 2D is
\begin{equation}
\langle \theta(\mathbf{r}) \theta(0) \rangle \sim \gamma'\log r \quad\Rightarrow\quad \langle A(\mathbf{r}) A(0) \rangle \sim 1/r^{\gamma'},
\end{equation}
i.e. algebraic long-range order with $\gamma' \sim \gamma/(\mathcal{S} A_0^2)$ and $A_0$ being the LC amplitude. This corresponds to effectively thermal behavior with an effective temperature set by the noise term $\zeta$ (which is proportional to $\gamma$).

\begin{figure}[t]
\begin{center}
\includegraphics[angle=0, width=1\columnwidth]{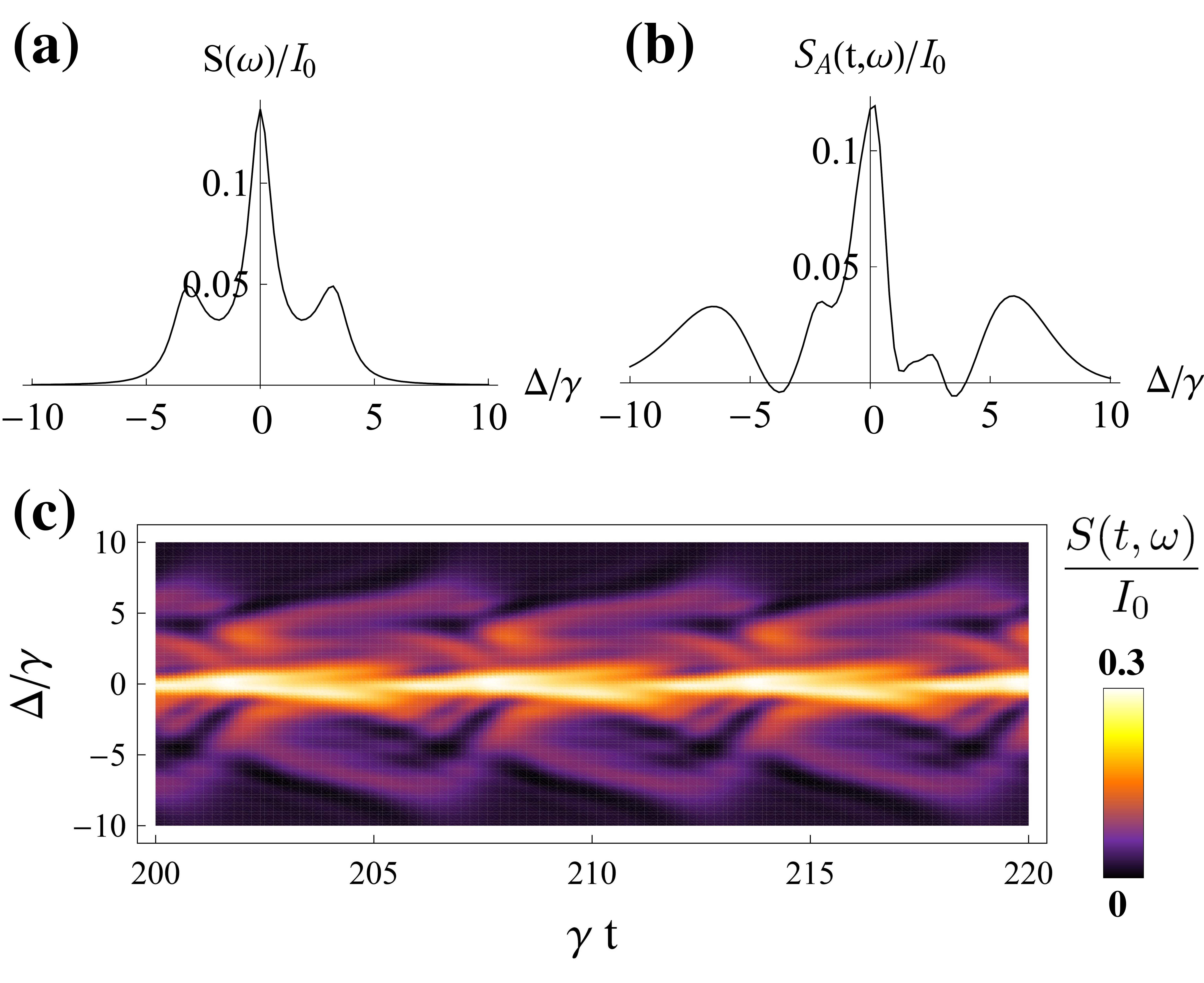}
\caption{(Color online) Power spectra as a function of detuning $\Delta$ with respect to the spin natural frequency in stationary and LC phases. (a) At $(\Omega,J_x,J_y,J_z)/\gamma=(0.6,-6,6,2)$, the system is in the time-independent $\text{U}_1$ phase and the corresponding power spectrum has a standard Mollow triplet structure. (b) The sublattice power spectrum for the LC phase at $(\Omega,J_x,J_y,J_z)/\gamma=(1,-7,6,2)$ and a particular time $t$. The asymmetry in $\Delta$ is caused by the complex-valued two-time correlation function in the LC phase. Both sublattice spectra have a time-dependence and they do not cancel each other. (c) The total dynamical power spectrum $S_A(t,\omega)+S_B(t,\omega)$ oscillates in time with the same periodicity of the LC. }
\label{fig_spectrum}
\end{center}
\end{figure}

\textit{Photon temporal correlation}.---The LC phase possesses novel dynamical features in the temporal fluctuations of spins, which can be seen in the emitted fluorescence. Each spin is coupled to the electromagnetic vacuum such that the time correlation of the fluorescence signal reflects that of the spin. For a conventional stationary state, like the $\text{U}_1$ phase, the two-time correlation function $\langle \sigma^+ (0) \sigma^-(\tau) \rangle$ has a damped oscillation, corresponding to a power spectrum with a three-peak structure [Fig.~\ref{fig_spectrum}(a)]. This is because the spins are collectively driven by an effective field and the dynamics are similar to those of a driven two-level system, so the photon power spectrum resembles a Mollow triplet \cite{scully97}.

However, the temporal dynamics are very different in the LC phase. A power spectrum is usually defined for a stationary state. Extending it to a time-dependent phase, the dynamical power spectrum is:
\begin{eqnarray}
\label{eq_spectrum}
S_L(t,\omega)=\frac{I_0}{\pi} \text{Re} \int_0^\infty d\tau \llangle \sigma^+_L (t) \sigma^-_L(t+\tau)\rrangle e^{i\omega \tau},
\end{eqnarray}
where the fluctuation is $\llangle \mathcal O_1 \mathcal O_2 \rrangle = \langle \mathcal O_1 \mathcal O_2\rangle - \langle \mathcal O_1\rangle \langle \mathcal O_2\rangle$, $I_0$ is the conventional power spectrum coefficient \cite{scully97} and $L$ denotes sublattice. For a stationary phase, only the time difference $\tau$ between the spin operators matters and the spectrum is independent of $t$. However, in the LC phase, $\langle \sigma^+_i (t) \sigma^-_i(t+\tau)\rangle \neq \langle \sigma^+_i (0) \sigma^-_i(\tau)\rangle$ and the spectrum becomes periodic in $t$. To calculate it, we focus on the parameter regime far away from the phase boundary so that spatial fluctuations are negligible and mean-field theory is valid (see Supplemental Material \cite{supple}).

Figure~\ref{fig_spectrum}(b) shows the power spectrum for the LC phase at a given time $t$ after the system reaches a steady state. The spectrum is \textit{asymmetric} in the detuning because $\langle \sigma^+ (t) \sigma^-(t+\tau)\rangle$ is no longer real due to the loss of time-translational invariance. Moreover, as illustrated in Fig.~\ref{fig_spectrum}(c), the power spectrum is periodic in time due to the oscillations. These dynamical properties are in sharp contrast to fluorescence signals in a stationary system [Fig.~\ref{fig_spectrum}(a)].

To probe the dynamical power spectrum, one can perform a time-resolved photon detection. The photon detection for a time $T$ is proportional to \cite{scully97}:
\begin{eqnarray}
\label{eq_photodetection}
P(T,\omega) \propto \frac{1}{T} \text{Re}\int_0^T dt  \int_0^\infty d\tau \llangle \sigma^+_L (t) \sigma^-_L(t+\tau) \rrangle e^{i\omega \tau},
\end{eqnarray}
where the average value is subtracted. A $T$-resolved photo-detection permits the measurement of $(T+\delta T)P(T+ \delta T,\omega)-T P(T,\omega) \approx \delta T\times S(T,\omega) $, when $\delta T$ is much shorter than the LC period.

In the case of quasi-long-range order or short-range order with a large correlation length, the power spectrum will become symmetric when averaged over long times. However, one expects that higher-order correlation functions, such as $\llangle \sigma_L(t) \sigma_L(t + \tau) \sigma_L(t') \sigma_L(t' + \tau) \rrangle$ (with $\tau \sim T$ and $|t - t'| \gg T$) will retain signatures of limit-cycle behavior, i.e., they will exhibit oscillations at a period $T$.

\textit{Conclusion.}---Driven-dissipative spin systems with anisotropic interactions have a rich nonequilibrium phase diagram; the most remarkable feature of this phase diagram is the presence of a limit-cycle phase. Although the limit cycle is inherently nonequilibrium (and thus exhibits an asymmetric power spectrum), this phase and its transitions can be understood in terms of familiar equilibrium concepts such as correlation lengths, Goldstone modes, and critical dimensions. We believe this work opens up a fruitful new interface between the condensed matter and nonlinear dynamics communities. Possible future directions include studying the criticality of spatiotemporal oscillation patterns \cite{liebchen14}, non-Markovian environments \cite{chan14}, and non-Hermitian spin systems \cite{lee14}.

This work was supported by the NSF through a grant to ITAMP. C.K.C. acknowledges the support from the Croucher Foundation.

\clearpage
\begin{widetext}

\begin{center}
\large{\bf Supplemental Material:\\ Limit cycle phase in driven-dissipative spin systems}\\
\vspace{14pt}
\normalsize{Ching-Kit Chan, Tony E. Lee, and Sarang Gopalakrishnan}
\end{center}
\vspace{14pt}

This supplement details the calculations of Gaussian theory and Floquet analysis for spatial correlations, a phenomenological derivation of the lower critical dimension and the photon temporal correlations in the LC phase.

\section{Equations of Motion and Mean-Field theory}

According to the master equation [Eq.~(\ref{eq_ME})] in the main text, the exact equations of motion for averages of one and two spin operators can be systematically written as:
\begin{eqnarray}
\label{eq_EOM1}
\frac{d}{dt}\langle \sigma^\alpha_n \rangle &=&  \epsilon^{\alpha \beta \gamma} \Omega^\beta \langle \sigma^\gamma_n \rangle   -  \Gamma^\alpha \langle \sigma^\alpha_n\rangle -  \gamma \delta^{\alpha z}+   \frac{1}{d} \sum_\mu  \epsilon^{\alpha \beta \gamma }  J^\beta \langle \sigma^\beta_{n+\mu} \sigma^\gamma_n \rangle, \\
\label{eq_EOM2}
\frac{d}{dt}\langle  \sigma^\alpha_n  \sigma^\beta_m \rangle &=&   \epsilon^{\alpha \gamma \delta} \Omega^\gamma \langle \sigma^\delta_n  \sigma^\beta_m \rangle  +  \epsilon^{\beta \gamma \delta} \Omega^\gamma \langle \sigma^\alpha_n  \sigma^\delta_m \rangle -  \left( \Gamma^\alpha  +\Gamma^\beta \right)\langle \sigma^\alpha_n \sigma^\beta_m \rangle -  \gamma \left[ \delta^{\alpha z} \langle \sigma^\beta_m \rangle +  \delta^{\beta z} \langle \sigma^\alpha_n\rangle \right] \nn \\
&&+   \frac{1}{d} \sum_\mu  \left [\epsilon^{\alpha \gamma \delta }  J^\gamma \langle \sigma^\delta_{n} \sigma^\gamma_{n+\mu} \sigma^\beta_m \rangle+ \epsilon^{\beta \gamma \delta }  J^\gamma \langle \sigma^\alpha_{n} \sigma^\gamma_{m+\mu} \sigma^\delta_m \rangle  \right],
\end{eqnarray}
where the subscript ($n, m$) and the superscript ($\alpha, \beta, \gamma, \delta$) denote the site and the spin index, respectively. Here, $\Gamma^x = \Gamma^y = \frac{\gamma}{2} $, $\Gamma^z = \gamma$ and $\mu = \pm \hat x, \pm \hat y, \pm \hat z$. The second equation only holds for $n \neq m$. We have also used the Levi-Civita symbol $\epsilon^{abc}$ to simplify the notations. Both $\delta^{\alpha\beta}$ and $\delta_{nm}$ are Kronecker deltas.

The sublattice mean-field decoupling of Eq.~(\ref{eq_EOM1}) (i.e. $\langle \sigma^\beta_{n+\mu} \sigma^\gamma_n \rangle \approx \langle \sigma^\beta_{B} \rangle \langle \sigma^\gamma_A \rangle = S^\beta_B S^\gamma_A$) results in a $6$-components self-consistent equation. A linear stability analysis of the steady state solution leads to the sublattice mean-field phase diagrams Fig.~\ref{fig_phase_diagram}(a) and Fig.~\ref{fig_phase_diagram_zoomout} in the main text. Note that there can be multisite instability such as spin-density-wave \cite{lee13_s} that goes beyond the sublattice mean-field theory. We have checked that and the spin-density-wave instability occurs within the $U_3$ phase and does not affect the LC.

\section{Gaussian spatial fluctuation}

Within the Gaussian framework, we define the two-point correlation functions and approximate higher order correlation functions as:
\begin{eqnarray}
\langle \sigma^\alpha_n \sigma^\beta_m\rangle &=& \bar \sigma^\alpha_n \bar\sigma^\beta_m + C^{\alpha,\beta}_{n,m}\ ,  \\
\langle \sigma^\alpha_n \sigma^\beta_m \sigma^\gamma_l \rangle & \approx & \bar\sigma^\alpha_n \bar\sigma^\beta_m \bar \sigma^\gamma_l + \bar \sigma^\alpha_n C^{\beta,\gamma}_{m,l} + \bar \sigma^\beta_m C^{\alpha,\gamma}_{n,l} + \bar \sigma^\gamma_l C^{\alpha,\beta}_{n,m}\ , \ \ \ \ \text{when  } n \neq m \neq l .
\label{eq_decoupling}
\end{eqnarray}
After a further introduce of sublattices $L= \pm 1$ (or A and B), the mean-field variable and the correlation function become $\bar \sigma^\alpha_{n} = \bar \sigma^\alpha_L $ and $C^{\alpha,\beta}_{n,m} = C^{\alpha,\beta}_{L,L',n-m}$. Applying these approximations, Eq.~(\ref{eq_EOM2}) becomes:
\begin{eqnarray}
\dot C^{\alpha,\beta}_{L,L',r} &=&  \epsilon^{\alpha \gamma \delta} \Omega^\gamma C^{\delta,\beta}_{L,L',r} +\epsilon^{\beta \gamma \delta} \Omega^\gamma C^{\alpha,\delta}_{L,L',r}  - \left( \Gamma^\alpha + \Gamma^\beta \right) C^{\alpha,\beta}_{L,L',r}   \nn \\
&& + \frac{1}{d}\sum_\mu \epsilon^{\alpha \gamma \delta} J^\gamma   \left\{ \bar \sigma^\delta_L C^{\gamma, \beta}_{-L,L',r+\mu} + \bar \sigma^\gamma_{-L} C^{\delta, \beta}_{L,L',r} + \delta_{-L,L'}\delta_{r+\mu,0}   \bar\sigma^\delta_L \left ( i\epsilon^{\gamma\beta\nu} \bar\sigma^{\nu}_{L'}+\delta^{\gamma\beta} - \bar\sigma^\gamma_{L'} \bar\sigma^\beta_{L'} - C^{\gamma,\beta}_{L',L',0} \right)   \right\} \nn \\
&& + \frac{1}{d}\sum_\mu \epsilon^{\beta \gamma \delta} J^\gamma   \left\{  \bar \sigma^\delta_{L'} C^{\alpha, \gamma}_{L,-L',r+\mu} + \bar \sigma^\gamma_{-L'} C^{\alpha, \delta}_{L,L',r} + \delta_{-L,L'}\delta_{r+\mu,0}  \bar\sigma^\delta_{L'} \left ( i\epsilon^{\alpha\gamma\nu} \bar \sigma^{\nu}_{L}+\delta^{\alpha\gamma} - \bar\sigma^\alpha_{L} \bar\sigma^\gamma_{L} -C^{\alpha,\gamma}_{L,L,0} \right)   \right\},
\label{eq_sublattice}
\end{eqnarray}
which is valid for $r \neq 0$. This equation can be understood in the followings. The first line comes from the driving and the decay contributions. In the second and third lines, the first two terms come from the Gaussian decoupling, and the third term accounts for situations with equal site indices. For example, when $r = -\mu$, terms like $\langle \sigma_0^a \sigma_{\mu}^b \sigma_{-r}^c \rangle$ only consist of spins from two different sites and shall not be decoupled. In this case, $\langle \sigma_0^a \sigma_{\mu}^b \sigma_{\mu}^c \rangle$ should be replaced by $\langle \sigma_0^a [i \epsilon^{bcd} \sigma_{\mu}^d +\delta^{bc}]\rangle$ instead.

To solve Eq.~(\ref{eq_sublattice}), we go to the momentum space. Defining $J^\alpha_k = J^\alpha \sum_\mu e^{i k \mu}$, the Fourier transform of Eq.~(\ref{eq_sublattice}) can be expressed as:
\begin{eqnarray}
\label{eq_Ck}
\dot {C}^{\alpha,\beta}_{L,L',k} &=&  A^{\alpha,\delta}_{L,L'',k} C^{\delta,\beta}_{L'',L',k} + C^{\alpha,\delta}_{L,L'',k} A^{\beta,\delta}_{L',L'',k} + D^{\alpha,\beta}_{L,L',k},
\end{eqnarray}
where A and D have the meanings of the drift and the diffusion matrix, respectively, with
\begin{eqnarray}
A^{\alpha,\delta}_{L,L',k} &=& \left(\epsilon^{\alpha\gamma\delta} \Omega^\gamma- \Gamma^\alpha \delta^{\alpha\delta} \right)\delta_{L,L'} +\frac{1}{d} \left( \epsilon^{\alpha\delta\gamma} J^\delta_k \bar\sigma^\gamma_L \delta_{L',-L} + \epsilon^{\alpha\gamma\delta} J^\gamma_0 \bar\sigma^\gamma_{-L} \delta_{L',L} \right),  \\
\label{eq_Dk}
D^{\alpha,\beta}_{L,L',k} &=& \frac{\delta_{-L,L'}}{d} \left[\epsilon^{\alpha\gamma\delta} J^\gamma_k \bar\sigma^\delta_L \left( i\epsilon^{\gamma\beta\nu} \bar\sigma^\nu_{L'}+\delta^{\gamma\beta} -\bar\sigma^\gamma_{L'} \bar\sigma^\beta_{L'} \right) + \epsilon^{\beta\gamma\delta} J^\gamma_k \bar\sigma^\delta_{L'} \left( i\epsilon^{\alpha\gamma\nu} \bar\sigma^\nu_{L}+\delta^{\alpha\gamma} -\bar\sigma^\alpha_{L} \bar\sigma^\gamma_{L} \right)\right] .
\end{eqnarray}
We have dropped qualitatively unimportant terms $\sim \mathcal O (1/d^2)$ in the D expression. For a stationary phase, the steady state correlation can be obtained by putting  $\dot {C}^{\alpha,\beta}_{L,L',k}=0$. However, in the LC phase, $\bar\sigma^\alpha_L(t)$, $A(t)$ and $D(t)$ are time-periodic, and one has to solve the $36$-components Eq.~(\ref{eq_Ck}) numerically.

\section{Floquet analysis for the spatial fluctuation}

We now extract the analytical properties of the solution of Eq.~(\ref{eq_Ck}), which is an inhomogeneous linear differential equation with time-periodic coefficients. It can be rewritten as:
\begin{eqnarray}
\label{eq_Ck2}
\dot {\vec  C}_k(t) = \boldsymbol M_k(t) \cdot \vec C_k(t) + \vec D_k(t),
\end{eqnarray}
where the vectors $\vec C_k(t)$ is the $36$-components correlation functions, $\boldsymbol M_k(t)$ and $\vec D_k(t)$ contains the drift and the diffusion coefficients, respectively. Let T be the period of $\boldsymbol M(t)$, $\vec D(t)$ and the LC. According to the inhomogeneous Floquet theory \cite{yakubovich75_s,slane11_s}, the solution of Eq.~(\ref{eq_Ck2}) can be expressed as follows:
\begin{eqnarray}
\label{eq_Ck3}
\vec  C_k(t) = \boldsymbol X_k(t) \left[  \vec C_k(0) + \int^t_0 d\tau \boldsymbol X_k^{-1}(\tau) \vec D_k(\tau) \right],
\end{eqnarray}
where $\boldsymbol X_k(t)$ is known as the principal fundamental matrix of Eq.~(\ref{eq_Ck2}). The Floquet theorem says that the fundamental matrix has the following form:
\begin{eqnarray}
\boldsymbol X_k(t) = \boldsymbol P_k(t) e^{\boldsymbol F_k t},
\end{eqnarray}
where $ \boldsymbol P_k(t)$ is some T-period matrix and $\boldsymbol F$ is some time-independent matrix. Note that $\boldsymbol X_k(t)$ is obtained from the homogeneous equation $\dot {\boldsymbol X}_k(t) = \boldsymbol M_k(t) \cdot \boldsymbol X_k(t)$, with an initial condition ${\boldsymbol X}_k(0) = \boldsymbol I$.

The stability of Eq.~(\ref{eq_Ck2}) is governed by the eigenvalues of $\boldsymbol X_k(t=T)$. These eigenvalues, denoted by $e^{\mu_i T}$, are known as the Floquet multipliers and $\mu_i$ are referred as the Floquet exponents. When $\mu_i < 0$, the result is stable; while if $\mu_i > 0$, the solution will grow exponentially in time. Figure~\ref{fig_floquet} presents the largest two Floquet exponents as a function of $k$ in the LC phase. In the long-wavelength limit, $\mu_1 \sim -k^2$ and $\mu_2 \sim -(k_0^2+k^2)$. Physically, $\mu_1$ and $\mu_2$ refer to the transverse and the longitudinal fluctuations, respectively, and $k_0 \sim \xi^{-1}$. Therefore, for finite $k$ values, the correlation function is stable and oscillatory in time [Fig.~\ref{fig_fluctuation}(d)]. The $k=0$ situation is special since the largest Floquet exponent is zero, corresponding to a Floquet instability.

\subsection{Correlation length, upper critical dimension and Ginzburg criterion}

With the help of Floquet exponents, we can obtain the analytical behavior of the correlation functions. One can show that for $t = n T$, Eq.~(\ref{eq_Ck3}) can be rewritten as \cite{slane11_s}:
\begin{eqnarray}
\vec  C_k(t=n T) = \boldsymbol X_k(T)^n \vec C_k(0) + \left[\boldsymbol X_k(T) + \boldsymbol X_k(T)^2 + ... +\boldsymbol X_k(T)^n \right] \int^T_0 d\tau \boldsymbol X_k^{-1}(\tau) \vec D_k(\tau).
\end{eqnarray}
For long time ($n \gg 1$), the above expression is dominated by the second inhomogeneous contribution. Diagonalizing $\boldsymbol X_k(T) = \boldsymbol U_k(T) \boldsymbol \Lambda_k(T) \boldsymbol U^{-1}_k(T)$ with $\left(\boldsymbol \Lambda_k(T)\right)_i = e^{\mu_i T}$, we can write in terms of the Floquet exponents:
\begin{eqnarray}
\label{eq_Ck4}
\left[\vec C_k(t=n T)\right]_i = \left[\boldsymbol U_k(T)\right]_{i,j} \left(\frac{e^{\mu_j T}-e^{\mu_j (n+1)T}}{1-e^{\mu_j T}}\right) \left[\boldsymbol U^{-1}_k(T) \int^T_0 d\tau \boldsymbol X_k^{-1}(\tau) \vec D_k(\tau)\right]_j,
\end{eqnarray}
Using the $k$ dependence of the dominant Floquet exponents, each stroboscopic correlation function in the long-wavelength limit can be expressed in this form:
\begin{eqnarray}
\label{eq_Ck5}
C_k(t=n T) \approx \frac{1-e^{-c n k^2}}{K_t k^2}+  \frac{1}{K_l\left(k^2+k_0^{2}\right)},
\end{eqnarray}
where $c$, $K_l$ and $K_t$ are some constant $\sim O(1)$. The first and second terms corresponds to transverse and longitudinal fluctuations, respectively. For finite time ($n$), the trasverse term accounts for the linear Floquet divergence of the correlation at $k=0$ [Fig.~\ref{fig_fluctuation}(b)] in the main text . $k_0^{-1}=\xi$ in the longitudinal term is the correlation length which diverges near the continuous phase transition [Fig.~\ref{fig_fluctuation}(d)].

\begin{figure}[t]
\begin{center}
\includegraphics[angle=0, width=.7\columnwidth]{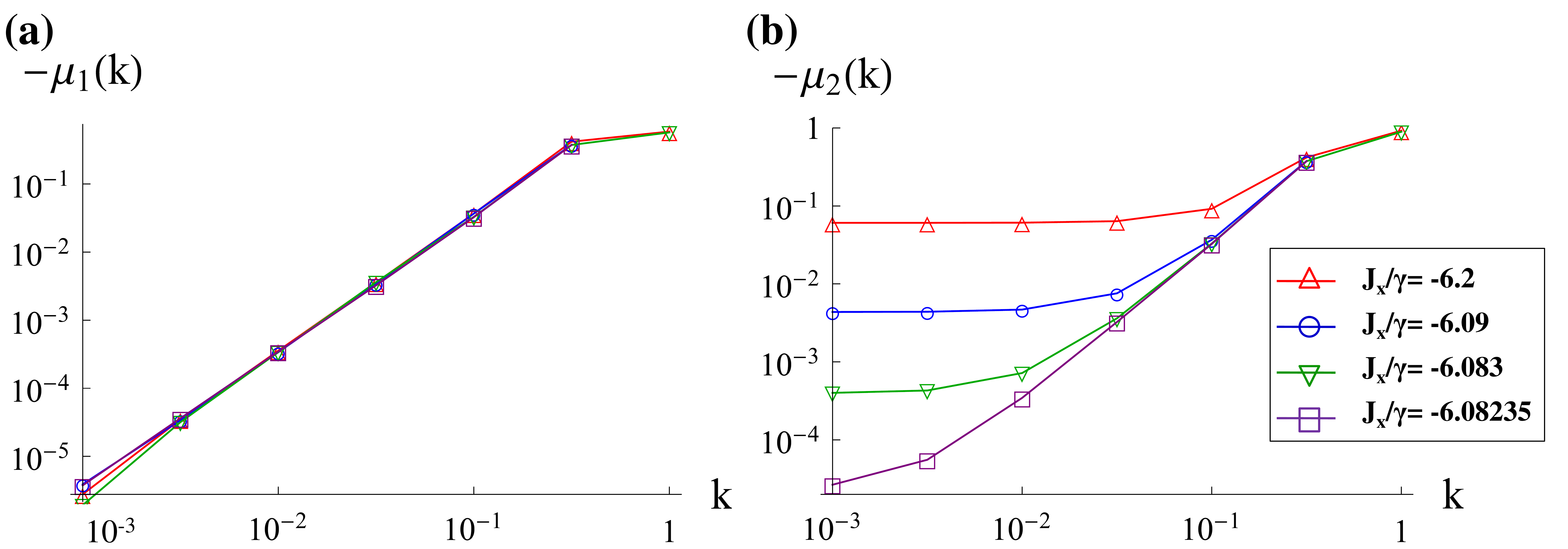}
\caption{(Color online) The first and second largest Floquet exponents $\mu_1$ and $\mu_2$ for the correlation functions $C(k,t)$ in the LC phase at $(\Omega,J_y,J_z)/\gamma=(0.6,6,2)$. $\mu_1 \propto -k^2$ and $\mu_2 \sim - (k^2+k_0^2)$ in the small $k$ limit. $k_0^{-1}$ is the inverse correlation length which diverges as $J_x/\gamma$ approaches the continuous phase transition.}
\label{fig_floquet}
\end{center}
\end{figure}

The correlation length near the LC-$\text{U}_1$ boundary diverges as $\xi \sim (\delta J_x/\gamma)^{-1/2}$. This property results in a Ginzburg criterion in $d$-dimension:
\begin{eqnarray}
\frac{C^{xx}(r\sim\xi,t)}{\left(\Delta S^x \right)^2 }\sim \frac{\xi^{4-d}}{K_l} \ll 1,
\end{eqnarray}
corresponding to a upper critical dimension of $4$ in the LC phase. Thus, despite its dynamical properties, the LC phase and its phase transition can still be characterized by a divergent correlation length and critical dimensions.


\section{From normal form to the lower critical dimension}

While the normal-form description (see main text) shows that the limit cycle has a Goldstone mode, and the Gaussian analysis above shows that its lower critical dimension is two, the link between these two observations is not explicit. Here, we address this gap. This is a standard calculation in the Keldysh formalism; we indicate the main steps, and refer the reader to the review article by Kamenev and Levchenko~\cite{kamenev09_s} (whose notation we follow) for details. We proceed by rewriting the normal-form equations using the Martin-Siggia-Rose approach~\cite{kamenev09_s,martin73_s}. We then transform to the rotating frame, and keep only the phase fluctuations of $A$ (i.e., $A(\mathbf{x},t) = A_0 \exp(i \theta(\mathbf{x},t))$). In terms of these phase fluctuations, the quadratic action (written in the classical-quantum basis~\cite{kamenev09_s}) is
\begin{equation}
S(\theta_c, \theta_q) = A_0^2 \left[\theta_q (\partial_t - \mathcal{S} \nabla^2) \theta_c + 2 i \zeta \theta_q^2 \right],
\end{equation}
where $\theta_{c,q}$ is the classical (quantum) phase field and $\zeta$ represents the Langevin noise amplitude. From this it is straightforward to compute the phase correlation function $\langle \theta_{\mathbf{k}} \theta_{-\mathbf{k}} \rangle$, as this is the Keldysh component of the $\theta$ Green's function computed from the action above. This gives the result quoted in the main text.

\section{Photon correlations}

To compute the photon power spectrum [Eq.~(\ref{eq_spectrum})] in the main text, we derive the exact equation of motion for the two-time correlation function from Eq.(\ref{eq_EOM1}):
\begin{eqnarray}
\label{eq_EOM3}
\frac{d}{d\tau}\langle \sigma^+_n(t) \sigma^\alpha_n(t+\tau) \rangle &=&  \epsilon^{\alpha \beta \gamma} \Omega^\beta \langle \sigma^+_n(t) \sigma^\gamma_n (t+\tau) \rangle   -  \Gamma^\alpha \langle\sigma^+_n(t)  \sigma^\alpha_n(t+\tau)\rangle -  \gamma \delta^{\alpha z} \langle \sigma^+_n(t) \rangle \nn \\
&&  + \frac{1}{d} \sum_\mu  \epsilon^{\alpha \beta \gamma }  J^\beta \langle \sigma^+_n(t) \sigma^\beta_{n+\mu}(t+\tau) \sigma^\gamma_n (t+\tau)\rangle .
\end{eqnarray}
We focus on the phase space where spatial fluctuation is negligible. In this case, the mean-field decoupling of the last interaction term (i.e. $\langle \sigma^+_n (t) \sigma^\beta_{n+\mu} (t+\tau)\sigma^\gamma_n (t+\tau)\rangle \approx \langle \sigma^\beta_{B}(t+\tau) \rangle \langle\sigma^+_A (t) \sigma^\gamma_A (t+\tau)\rangle$) results in a differential equation with time-dependent coefficients. We numerically solve this equation to obtain the power spectrum. The fact that the Eq.~(\ref{eq_EOM3}) shares the same structure as Eq.~(\ref{eq_EOM1}) and is usually referred as the quantum regression theorem \cite{scully97_s}.

\end{widetext}


\begin{thebibliography}{99}
\bibitem{note1} In equilibrium, the noise and dissipation would be related by the fluctuation-dissipation theorem; however, in a general nonequilibrium setting they are independent.
\end{thebibliography}
\end{document}